\documentclass[aps,prb,superscriptaddress,sd,amsmath,reprint,amssymb]{revtex4-1}
\usepackage[T1]{fontenc}

\usepackage[utf8]{inputenc}
\usepackage{graphicx}
\usepackage{dcolumn}
\usepackage{bm}
\usepackage{color}
\usepackage{eurosym}
\usepackage{hyperref}

\begin{document}
\title{Dynamics of Li deposition on epitaxial graphene/Ru(0001) islands}

\author{J.E. Prieto}
\email{jprieto@iqfr.csic.es}
\affiliation{Instituto de Qu\'{\i}mica F\'{\i}sica ``Rocasolano'', CSIC, Madrid E-28006, Spain}

\author{M.A. González-Barrio}
\affiliation{Dpto. de F\'{\i}sica de Materiales, Universidad Complutense de Madrid, 28040 Madrid, Spain}

\author{E. García-Martín}
\affiliation{Instituto de Qu\'{\i}mica F\'{\i}sica ``Rocasolano'', CSIC, Madrid E-28006, Spain}

\author{G.D. Soria}
\affiliation{Instituto de Qu\'{\i}mica F\'{\i}sica ``Rocasolano'', CSIC, Madrid E-28006, Spain}

\author{L. Morales de la Garza}
\affiliation{Centro de Nanociencias y Nanotecnología, Universidad Nacional Autónoma de México, Ensenada, BC, México}

\author{J. de la Figuera}
\affiliation{Instituto de Qu\'{\i}mica F\'{\i}sica ``Rocasolano'', CSIC, Madrid E-28006, Spain}

\date{\today}

\begin{abstract}
Li metal has been deposited on the surface of a Ru(0001) single crystal containing patches of monolayer-thick epitaxial graphene islands. The use of low-energy electron microscopy and diffraction allowed us to {\em in situ} monitor the process by measuring the local work function as well as to study the system in real and reciprocal space, comparing the changes taking place on the graphene with those on the bare Ru(0001) surface. It is found that Li deposition decreases the work function of the graphene islands but to a much smaller degree than of the Ru(0001) surface, as corresponds to its intercalation below the graphene overlayer. Finally, the diffusion process of Li out of the graphene islands has been monitored by photoelectron microscopy using a visible-light laser. 
\end{abstract}
\maketitle

\section{Introduction}
Graphene is a fascinating material with extraordinary mechanical, electronic and optical properties and actual or promising applications in many fields; to mention a few: microelectronics, spintronics, solar cells, batteries or water filters. Furthermore, intercalation of atoms into graphite and other layered materials\cite{DaukiyaPSS2019} has long been known to produce compounds with new electrical and chemical properties as compared with those of the original material. For example, intercalation into bulk graphite leads to compounds with different superconducting\cite{Smith2015} properties. Also lithium-intercalated graphite has found application as the anode material in rechargeable batteries \cite{DahnScience1995,TarasconNature2001,AricoNatMat2005,AsenbauerSusEn2020}. While the bulk material has been widely studied \cite{WertheimSSC1980,EberhardtPRL1980,FausterPRL1983,HolzwarthPRB1978,HolzwarthPRB1984,HatwigsenPRB1997,CsanyiNatPhys2005}, the manufacturing of Li-graphene represents an important step towards nanometer-sized ion batteries with high charge-discharge ratios \cite{AricoNatMat2005}.
Useful Li ion cycling has been shown in thin graphene/Ni films \cite{RadhakrishnanJElecSoc2012,DavidACSInterf2013} and single layer graphene has been proposed as a protective layer against corrosion for battery components \cite{YaoJACS2012}. This opens the path towards fabricating flexible electrodes for micro- or nano-scale batteries. Li-ion batteries based on graphene and Si have also been demonstrated with outstanding performances\cite{AndronicoLaptop}.

Epitaxial graphene, particularly on metal substrates\cite{WintterlinSS2009} has been shown to allow intercalation \cite{VirojanadaraPRB2010,EmtsevPRB2011,GierzPRB2010,YagyuAPL2014,ZhangNanotech2017}. However, the process through pristine graphene layers is governed by native defects and highly perfect layers can allow intercalation only in domains of reduced the lateral scale. Effective intercalation has been attributed to the presence of cracks and defects in the layers caused by initial deposition of the intercalant \cite{VirojanadaraPRB2010}. In addition, increasing the defect density by plasma etching has been shown to successfully lead to large-area intercalation\cite{BriggsNatMat2020}. It is therefore of great interest to study the process of Li incorporation into single-layer graphene with high spatial resolution in order to gain insight into the role of defects and edges in the intercalation process. To such end we use epitaxial graphene on Ru(0001), which can be considered \cite{SutterNatMat2008} a model system for the growth and study of high quality graphene. Detailed studies are available on the growth \cite{McCartyCarbon2009,CuiPCCP2010} as well as on the influence of interactions with the substrate\cite{BorcaNJP2010}. Intercalation of many types of atoms has also been explored in this system \cite{HuangAPL2011}, including selected cases such as oxygen, which can be used to decouple the graphene layer from the Ru substrate\cite{UlstrupSS2018}. In this work, we thus set out to locally study the process of Li incorporation into highly perfect, epitaxial monolayer graphene islands on Ru(0001) using low-energy electron microscopy (LEEM) and diffraction (LEED).

\section{Experimental}

Experiments were carried out in an ultra-high vacuum chamber (base pressure: 5.0$\times$10$^{-10}$ mbar) equipped with a low-energy electron microscope\cite{BauerBook2014} Elmitec LEEM III. In addition to other facilities not used in the present work, the system contains a SAES-getter type Li source and a leak valve allowing a precise dosing of molecular oxygen gas. The substrate used was a Ru(0001) single crystal. The standard procedure for Ru(0001) surface cleaning consists of cycles of flashing up to about 1500~K and exposing to molecular oxygen at pressures in the range of 10$^{-7}$ mbar and temperatures of about 900~K. Annealing at these temperatures segregates carbon dissolved from the bulk to the surface, which reacts with oxygen and desorbs. If no oxygen is present and if the near-surface region is not yet fully depleted of carbon, it accumulates on the surface in the form of monolayer patches of graphene. The process can be followed in real time by LEEM imaging\cite{McCartyCarbon2009}. 

\section{Results and Discussion}
\begin{figure}
    \centering
    \includegraphics[width=0.5\textwidth]{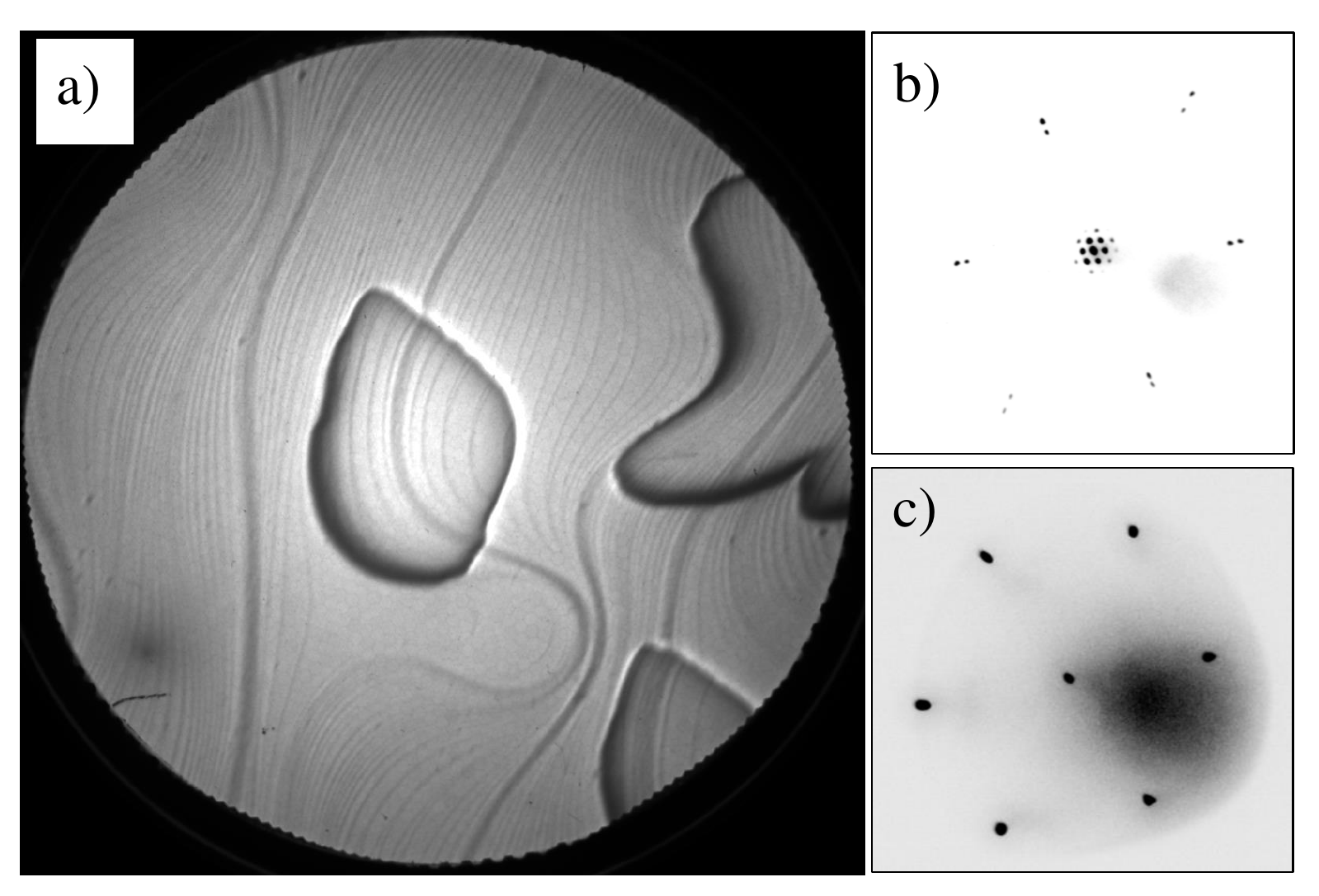}
    \caption{(a): LEEM image of a Ru(0001) surface containing three graphene islands, 10~$\mu$m. SV = 19.2~V; LEED patterns of: (b): LEED pattern of a graphene island, E = 46.7~eV, showing a moiré superstructure; (c) (1$\times$1) LEED pattern of the clean Ru(0001) surface, E = 45.0~eV. The diffuse intensity is due to secondary electrons.}
    \label{fig:graphene}
\end{figure}

Graphene islands can be grown from carbon dissolved in the bulk of a Ru crystal by annealing at temperatures of 900~K, as has been known for a long time\cite{SutterNatMat2008}. An example is shown in Fig.~\ref{fig:graphene}a. Here the Ru(0001) surface is imaged, with its monoatomic steps separating atomically flat terraces. Additionally, large carbon islands, with lateral sizes of up to several microns can be observed. The carbon layer is transparent enough to the electron beam for the substrate steps underneath to be clearly observed.

That the islands are indeed composed of carbon in the form of graphene is shown by their characteristic LEED pattern on Ru(0001)\cite{MoritzPRL2010}. By inserting an illumination aperture into the incoming electron beam and thus restricting the illuminated area to one such island, the diffraction pattern corresponding to the island’s surface can be recorded, performing in this way Selected Area Diffraction,
as shown in Fig.~\ref{fig:graphene}b. This is a {\em moiré} pattern formed due to the coincidence of the lattices of the graphene overlayer (lattice constant: 0.246~nm) and the Ru(0001) surface (lateral lattice parameter: 0.271 nm). The (1$\times$1) pattern (Fig.~\ref{fig:graphene}c) of the latter can be observed on the clean regions uncovered by the graphene.

\begin{figure}
    \centering
    \includegraphics[width=0.5\textwidth]{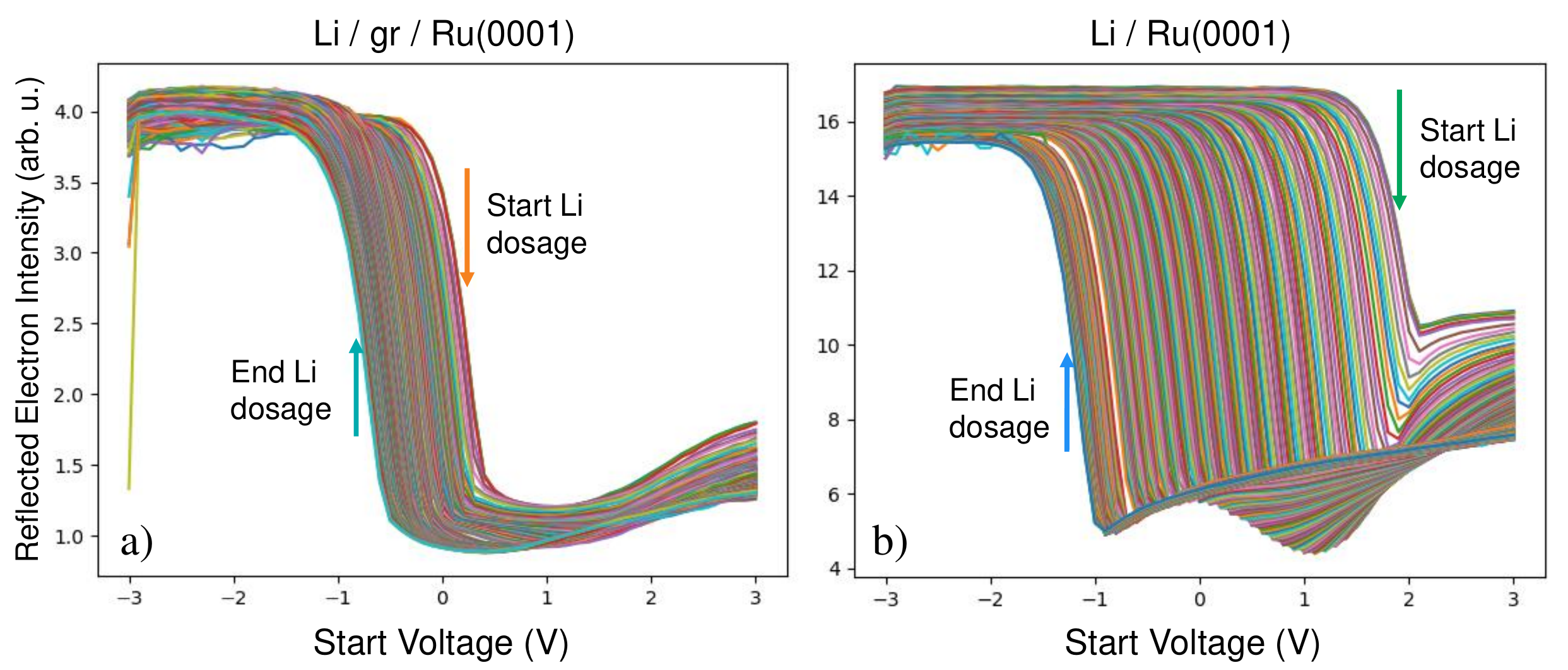}    \caption{Sequence of electron reflectivity vs. start voltage continuously recorded during Li deposition on top of (a) a graphene island, (b) on a clean Ru(0001) region. Deposition time is 3~h for (a) and 1~h 45~min for (b).}
    \label{fig:scanSV}
\end{figure}

Li was deposited onto Ru(0001) surfaces containing epitaxial graphene islands. The work function was continuously monitored during Li dosing. For this purpose, LEEM images were recorded as a function of the start voltage (SV), defined as the potential difference between the cathode and the sample. The SV was swept typically between -3.0~V and 3.0~V repeatedly during Li evaporation. The reflected electron intensity was integrated inside a user-defined window that can be placed either on a graphene island or on the clean Ru(0001) surface. Typical resulting reflectivity vs. SV curves are shown in Fig.~\ref{fig:scanSV}. Incoming electrons with kinetic energies lower than the sample’s work function cannot penetrate into the solid and are completely reflected at the surface. Therefore, upon lowering the SV, a transition to a nearly complete reflectivity (mirror mode\cite{BauerBook2014}) takes place when the SV equals the difference of work functions of sample and cathode. This transition can be used to monitor the sample’s work function evolution.
In this work, its precise value has been taken as the SV at which the derivative of the reflectivity curve attains its maximum absolute value. 

Figure~\ref{fig:scanSV} displays two sets of reflectivity vs. SV curves recorded during Li deposition: (a) On a graphene island, for a deposition time of 3~h, and (b) on a clean Ru(0001) region, for a deposition time of 1~h 45~min. Although there is in both cases a noticeable decrease of the work function upon Li-deposition, it is clear that the effect is less pronounced in the graphene / Ru system, as will be commented below.

\begin{figure}
    \centering
    \includegraphics[width=0.45\textwidth]{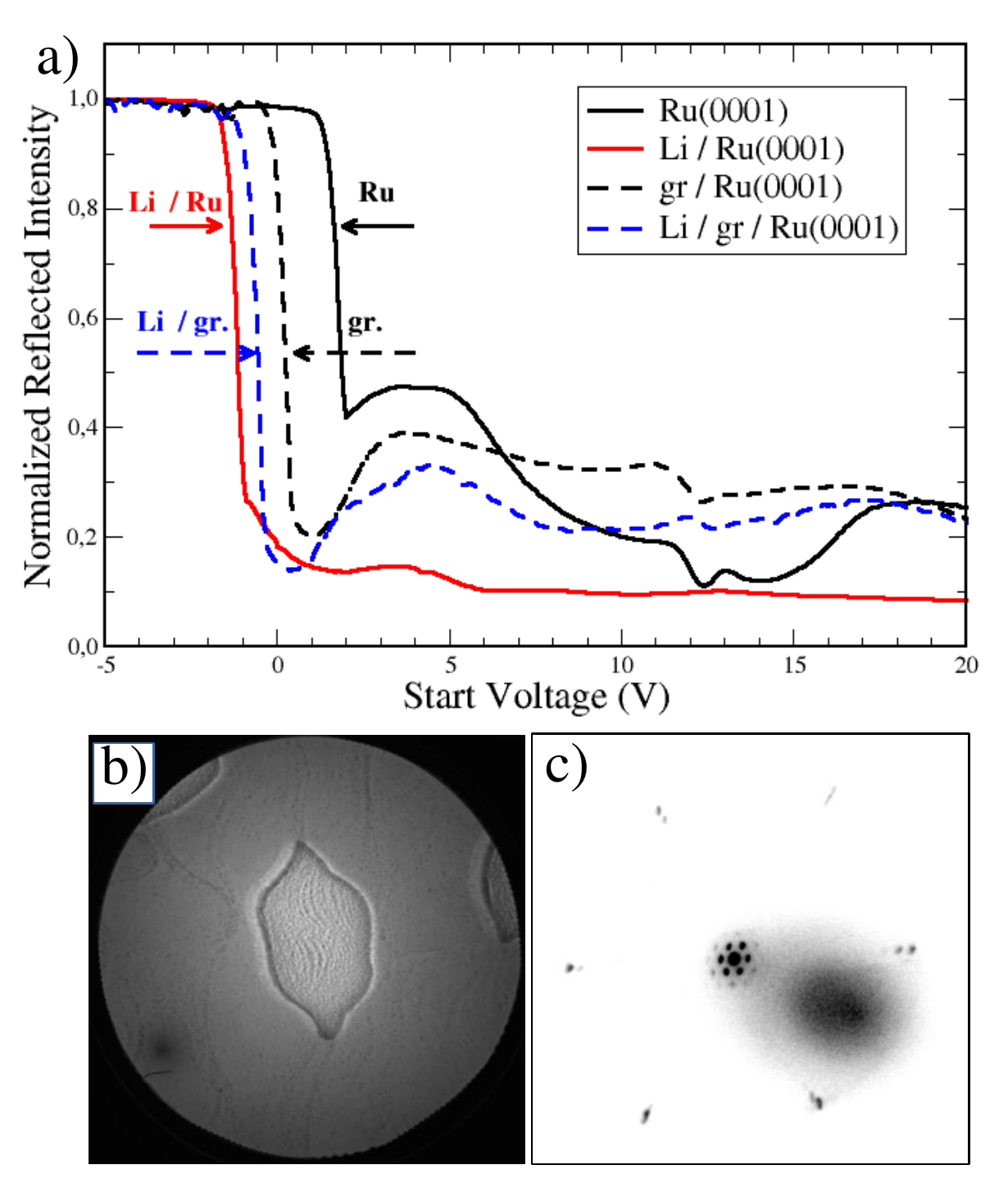}
    \caption{a): Electron reflectivity vs. start voltage measured for the clean and Li-covered Ru(0001) surfaces (solid lines) and for the clean and Li-covered graphene/Ru(0001) islands (coloured dashed lines). Deposition time is 3~h and the coverage amounts to about 0.5~ML; b): Li-covered graphene island at a start voltage of -0.46~V (field of view is 10~$\mu$m); c): LEED pattern of a Li-covered graphene island at an electron energy of 45~eV.}
    \label{fig:InitialFinal}
\end{figure}

After the completion of each experiment of Li deposition, a new reflectivity curve was recorded in a wider energy range (-5.0~eV - 20.0~eV). One example is shown in Figure~\ref{fig:InitialFinal} with a dashed blue line (corresponding to the experiment of Fig.~\ref{fig:scanSV}a), together with the corresponding curve on the graphene island before deposition for comparison (dashed black line). The total amount of Li deposited is close to that producing the maximum reduction of the work function in the system, as will be shown below. 
Data have been normalized to the maximal electron reflected intensity. The decrease of the work function of the graphene island, measured about 5~min after finishing the Li deposition (Fig.~\ref{fig:InitialFinal}) amounts to 0.8~eV.
One point to be mentioned is that, as deduced from the first and last curves recorded in real time during deposition (Fig.~\ref{fig:scanSV}), the decrease in work function amounts to 1.0~eV. 
This means that the work function of the Li/graphene/Ru(0001) system slightly increases in the time following the end of the Li deposition, showing that there is a measurable evolution at room temperature of the Li-covered surface. 

The local character of LEEM microscopy allows us to compare with the evolution of the clean Ru(0001) surface upon Li deposition. For this purpose, the corresponding reflectivity curves of the clean and Li-covered Ru(0001) surfaces, after deposition of the same amount of Li, are included in Figure~\ref{fig:InitialFinal} with solid lines. For Ru(0001), the reduction of the work function upon Li deposition amounts to 3.3~eV, close to the maximal reduction in this system, in agreement with published results\cite{Bromberger2002}, compared to the 0.8~eV found for a similar coverage on the epitaxial graphene/Ru(0001) islands (Fig.~\ref{fig:InitialFinal}a, dashed lines).

After Li deposition, the typical aspect of the graphene/Ru(0001) islands is shown in Fig.~\ref{fig:InitialFinal}b). 
A somewhat granular structure can be appreciated inside the islands. In addition, they display a LEED pattern, shown in Fig.~\ref{fig:InitialFinal}c), with a similar moiré structure as before deposition (Figure~\ref{fig:graphene}b), now with higher background intensity. 
Thus, no new superstructure is formed, either with respect to the substrate Ru(0001) surface or to the graphene overlayer, upon adsorption or incorporation of Li in the graphene islands.

\begin{figure}
    \centering
    \includegraphics[width=0.45\textwidth]{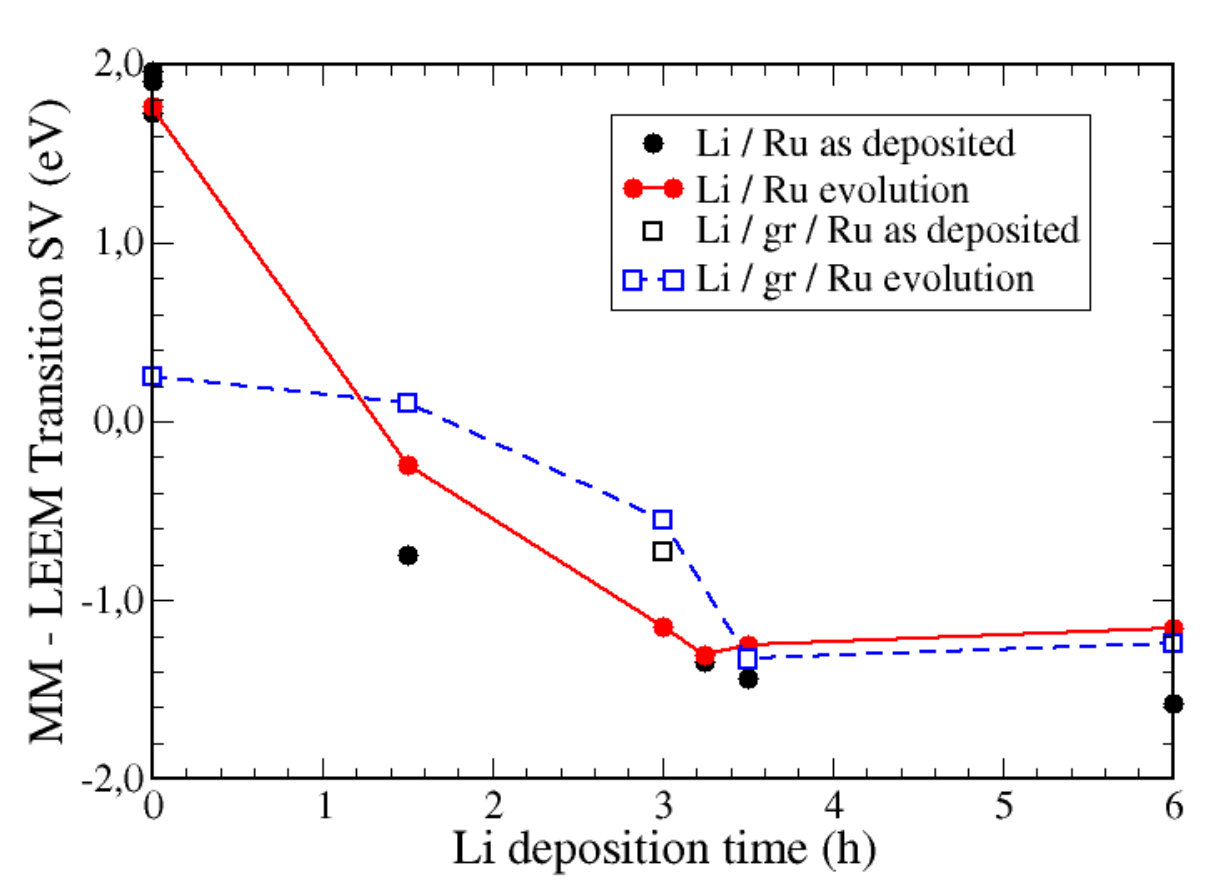}
    \caption{Work function decrease, as measured by the Mirror-Mode to LEEM transition start voltage for Li deposition on clean Ru(0001) and on graphene/Ru(0001). Black symbols correspond to the data obtained in real time during Li deposition, while the colored ones correspond to the reflectivity curves several minutes after finishing deposition, as explained in the text. Red and blue symbols refer to clean Ru(0001) and graphene/Ru(0001), respectively.}
    \label{fig:compilation}
\end{figure}



A compilation of results for the evolution of the work function of both the graphene islands and the uncovered Ru(0001) surface as a function of the Li deposition time is shown in Fig.~\ref{fig:compilation}. It can be seen that in the case of the Ru(0001) surface, the work function initially decreases rapidly with increasing Li coverage, reaching a maximum reduction of about 3.3~eV for the clean Ru surface (corresponding to a coverage of 0.3~ML-0.5~ML\cite{Bromberger2002}).
On the other hand, the work function of the graphene/Ru islands decreases initially very slowly and then a more rapid decrease takes place reaching a maximal reduction of about 1.5~eV. This is achieved for a similar coverage as the one which gives the maximal reduction of work function for Li/Ru(0001). 
For higher Li coverages, the work function evolution shows a slight increase (about 0.2~eV for the clean Ru surface, about 0.1~eV for graphene/Ru as deduced from the measurements performed several minutes after Li deposition). The behavior for Li/Ru(0001) is the canonical one observed for alkali-atom deposition on transition-metal surfaces \cite{ZangwillBook}. The strong initial decrease is due to the formation of a surface dipole layer as a consequence of the large electron transfer from the highly electronegative alkali atoms to the transition metal. This process reaches its maximum for an alkali coverage of the order of 0.5~ML. For higher coverages a small increase of the work function takes place due to depolarizing effect of neighbour dipoles (Topping\cite{Topping1927} model) and possibly also to a decrease in the electronic polarizability of the charge cloud due to electron transfer \cite{VerhoefSS1997}. 

An aspect which needs to be addressed is the effect on the measured work functions of the local electric fields arising from the coexistence of regions (patches) of different work functions as the Li coverage increases. At the beginning of deposition, Li atoms are expected to be randomly distributed on the surface, well separated from each other due to the electrostatic repulsion of dipoles. 
These conditions correspond to the small-patch regime, in which the measured work function gives a surface-averaged value allowing the determination of the induced surface dipole\cite{Bundaleski2013}. As coverage increases, condensation of adsorbate islands sets in and thus a coexistence of patches of surface covered by dilute Li atoms and surface covered by condensed Li atoms is expected, the latter with higher work function due to the depolarizing field of nearby dipoles. At some coverage, we must thus enter into the large-patch regime, favoured by the high extraction field produced by the 10~kV applied to the objective lens of the LEEM, of order 10$^4$~V/mm. 
This can be associated with the granular aspect of the graphene islands, as well as the Ru substrate, after Li deposition (Fig.~\ref{fig:InitialFinal}b).
In this regime the 
emission or absorption of electrons 
is dominated by the low work function patches independently of their size, so that the dependence of the measured work function on coverage tends to become weaker and a smooth decrease approaching saturation is expected. These considerations apply to the Li-covered regions both of the Ru(0001) substrate and of the graphene/Ru(0001) islands.  
The time dependence found in the system, i.e., the increase of work function measured following Li deposition can be ascribed to the rearrangement of Li, favoured by its high mobility, from the dilute-dipole to the compact island phases with a higher work function.

\begin{figure}
    \centering
    \includegraphics[width=0.5\textwidth]{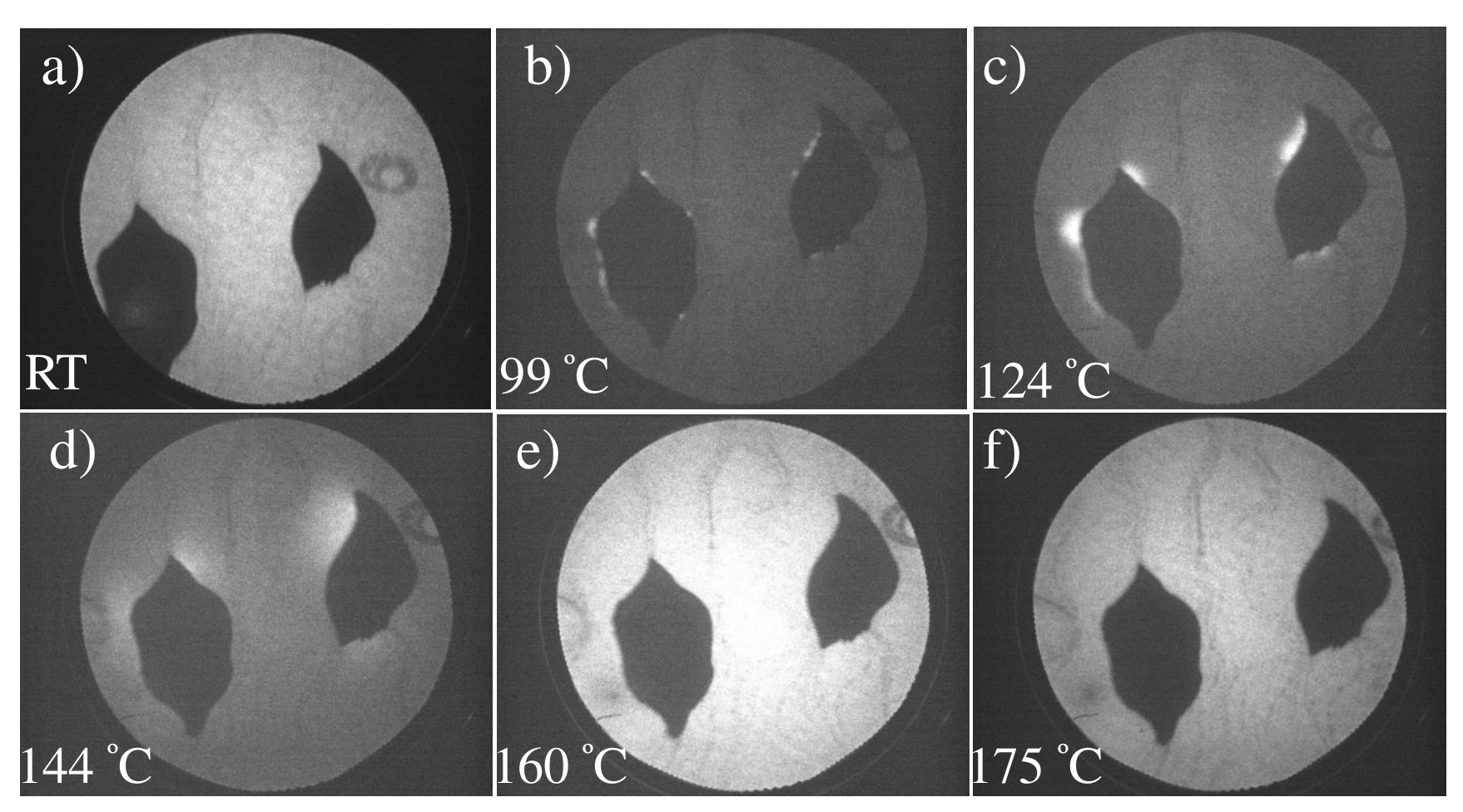}
    \caption{PEEM images using a blue laser ($h\nu$~= 3.06~eV) of a Ru(0001) surface containing two graphene islands. The size of the images is 20~$\mu$m. Images were taken at the quoted temperatures. Except for the first, taken at room temperature, images have been recorded using the same settings so that intensities can be directly compared.}
    \label{fig:PEEM}
\end{figure}

The maximum work function reduction upon Li deposition is a factor of 2 smaller for graphene/Ru than for the clean Ru(0001) surface (1.5~eV vs. 3.3~eV). 
Considering the electronegativities of the atoms involved (difference in Mulliken’s electronegativity for Li-C is 6.8~eV, compared to 5.4~eV for Li-Ru\cite{MullikenJCP1934,HinzeJACS1962}), we would expect a higher electron transfer for Li/C than for Li/Ru and therefore, a larger reduction of the work function for the Li/graphene/Ru(0001) surface than for Li/Ru(0001), if Li is residing on the surface of the graphene islands.
It could be even larger considering that a charge transfer from the C atoms to the Ru substrate has already taken place, as shown by the work function reduction of gr/Ru(0001) by about 1.6~eV respect to Ru(0001), leaving the graphene surface positively charged.
However, the opposite is found. This strongly suggests that the Li atoms are not simply sitting on top of the graphene/Ru(0001) surface. Such would be the situation if the Li atoms were being intercalated at the graphene/Ru interface upon deposition already at room temperature. This is in agreement with calculations indicating that Li cannot reside on the surface of defect-free graphene\cite{Lee_Persson2012,Dimakis2019}.
The picture of Li atoms falling on top of the graphene islands, diffusing to the edges and from there onto the Ru terraces is consistent with the sharp decrease of the work function that takes place only after an appreciable Li coverage; the intercalation would then significantly proceed only after the Ru(0001) terraces have been covered by 0.3~ML - 0.5~ML Li and by a mechanism of diffusion following a concentration gradient across the island edges into the interior of the islands.


The maximal reduction of the work function of Ru(0001) upon Li adsorption amounts to 3.3~eV. Since the work function of the clean surface is 5.5~eV\cite{WandeltSS1981}, the final value, 2.3~eV, is low enough to allow photoemission of electrons by visible light. Illuminating with a violet laser (photon energy: 3.06~eV) and collecting the produced photoelectrons with the image column of the microscope allows us to image the Li-covered Ru(0001) with its graphene islands, performing in this way photoemission electron microscopy (PEEM). Examples of PEEM images of a Li-covered Ru(0001) surface containing two graphene islands can be seen in Figure~\ref{fig:PEEM}.

A final experiment in PEEM was performed in order to gain insight into the presence of Li in the graphene islands. After deposition of about 0.5~ML of Li, the surface was continuously imaged in PEEM mode while raising the temperature. A selection of images of the process at different temperatures is shown in Figure~\ref{fig:PEEM}. In the image at 99$^\circ$~C (b), small bright spots can be seen in some regions at the edges of the graphene islands (preferentially at the corners) which were not visible in the images at room temperature. As the temperature is raised, these regions first become brighter and then appear to grow, i.e. the regions of higher photoemission yield become wider and
spread from the graphene islands away, while their intensity reduces as they expand across the Ru(0001) terraces, until they finally occupy the complete surface uncovered by the graphene islands. This suggests that temperature promotes the flow of Li out of the islands at their edges, mainly at the corner points (and suggests these can be also preferred places for Li intercalation into the islands) and its diffusion onto the Ru terraces, increasing the amount of Li adsorbed there and promoting a further reduction of the work function and an increase in photoelectron intensity. This is a further support for the fact that Li atoms were originally intercalated below the graphene layer following deposition. The Li density there can be higher than on the Ru(0001) surface after deposition because of the higher affinity of Li to C than to Ru, as quantified by the higher electronegativity difference between Li and~C. Thermal activation is then likely to promote diffusion of Li out of the graphene islands onto the Ru(0001) terraces following the concentration gradient. This process sets in at temperatures around 100$^{\circ}$~C, from which an activation barrier for Li diffusion of the order of 40~meV can be deduced. This small value is reasonable and explains the time evolution of the work function after finishing Li deposition (see Fig.~\ref{fig:compilation}) and the high mobility of Li atoms. It compares also well with the 46~meV calculated for the similar case of Na diffusion on Ru(0001)\cite{Raghavan2021}.

\section{Conclusions}

Summarizing, we have grown submonolayer islands of epitaxial graphene on Ru(0001) and have monitored their morphology, structure and electronic properties (work function) by means of LEEM in real time during Li deposition, comparing with the evolution of the clean Ru(0001) surface. Li reduces the work function of graphene/Ru(0001) by as much as 1.5~eV, significantly less than the 3.3~eV found for Ru(0001). Intercalation takes place at room temperature without forming any new superstructure. This behavior is confirmed by observing the temperature evolution with visible-light PEEM, taking advantage of the work function reduction in the system. Temperature promotes the diffusion of Li atoms out of the graphene islands preferentially at their corners and their expansion onto the Ru(0001) terraces.

\section{Aknowledgments}

This work is supported by grant RTI2018-095303-B-C51, funded by MCIN\-/AEI/10.13039/501100011033 and by “ERDF A way of making Europe”, by grant PID2020-117024GB-C43 (Ministerio de Ciencia e Innovación) and by grant S2018-NMT-4321, funded by the Regional Government of Madrid and by “ERDF A way of making Europe”.
L.M.G. acknowledges a sabbatical grant from DGAPA-UNAM.

\section*{References}
\bibliographystyle{h-physrev.bst}
\bibliography{Li_Gr}

\begin{thebibliography}{10}

\bibitem{DaukiyaPSS2019}
L.~Daukiya {\em et~al.},
\newblock Progress in Surface Science {\bfseries 94}, 1 (2019).

\bibitem{Smith2015}
R.~P. Smith {\em et~al.},
\newblock Physica C: Superconductivity and its Applications {\bfseries 514}, 50
  (2015).

\bibitem{DahnScience1995}
J.~R. Dahn, T.~Zheng, Y.~Liu, and J.~S. Xue,
\newblock Science {\bfseries 270}, 590 (1995).

\bibitem{TarasconNature2001}
J.-M. Tarascon and M.~Armand,
\newblock Nature {\bfseries 414}, 359 (2001).

\bibitem{AricoNatMat2005}
A.~S. Aricò, P.~Bruce, B.~Scrosati, J.-M. Tarascon, and W.~van Schalkwijk,
\newblock Nature Materials {\bfseries 4}, 366 (2005).

\bibitem{AsenbauerSusEn2020}
J.~Asenbauer {\em et~al.},
\newblock Sustainable Energy \& Fuels {\bfseries 4}, 5387 (2020).

\bibitem{WertheimSSC1980}
G.~K. Wertheim, P.~T. T.~M. Van~Attekum, and S.~Basu,
\newblock Solid State Communications {\bfseries 33}, 1127 (1980).

\bibitem{EberhardtPRL1980}
W.~Eberhardt, I.~T. McGovern, E.~W. Plummer, and J.~E. Fisher,
\newblock Physical Review Letters {\bfseries 44}, 200 (1980).

\bibitem{FausterPRL1983}
T.~Fauster, F.~J. Himpsel, J.~E. Fischer, and E.~W. Plummer,
\newblock Physical Review Letters {\bfseries 51}, 430 (1983).

\bibitem{HolzwarthPRB1978}
N.~A.~W. Holzwarth, S.~Rabii, and L.~A. Girifalco,
\newblock Phys. Rev. B {\bfseries 18}, 5190 (1978).

\bibitem{HolzwarthPRB1984}
N.~A.~W. Holzwarth, S.~G. Louie, and S.~Rabii,
\newblock Phys. Rev. B {\bfseries 30}, 2219 (1984).

\bibitem{HatwigsenPRB1997}
C.~Hartwigsen, W.~Witschel, and E.~Spohr,
\newblock Phys. Rev. B {\bfseries 55}, 4953 (1997).

\bibitem{CsanyiNatPhys2005}
G.~Csányi, P.~B. Littlewood, A.~H. Nevidomskyy, C.~J. Pickard, and B.~D.
  Simons,
\newblock Nature Physics {\bfseries 1}, 42 (2005).

\bibitem{RadhakrishnanJElecSoc2012}
G.~Radhakrishnan, J.~D. Cardema, P.~M. Adams, H.~I. Kim, and B.~Foran,
\newblock Journal of The Electrochemical Society {\bfseries 159}, A752 (2012).

\bibitem{DavidACSInterf2013}
L.~David {\em et~al.},
\newblock ACS Applied Materials \& Interfaces {\bfseries 5}, 546 (2013).

\bibitem{YaoJACS2012}
F.~Yao {\em et~al.},
\newblock Journal of the American Chemical Society {\bfseries 134}, 8646
  (2012).

\bibitem{AndronicoLaptop}
M.~A. Editor,
\newblock LaptopMag  (2014).

\bibitem{WintterlinSS2009}
J.~Wintterlin and M.-L. Bocquet,
\newblock Surface Science {\bfseries 603}, 1841 (2009).

\bibitem{VirojanadaraPRB2010}
C.~Virojanadara, S.~Watcharinyanon, A.~A. Zakharov, and L.~I. Johansson,
\newblock Physical Review B {\bfseries 82}, 205402 (2010).

\bibitem{EmtsevPRB2011}
K.~V. Emtsev, A.~A. Zakharov, C.~Coletti, S.~Forti, and U.~Starke,
\newblock Physical Review B {\bfseries 84}, 125423 (2011).

\bibitem{GierzPRB2010}
I.~Gierz {\em et~al.},
\newblock Physical Review B {\bfseries 81}, 235408 (2010).

\bibitem{YagyuAPL2014}
K.~Yagyu {\em et~al.},
\newblock Applied Physics Letters {\bfseries 104}, 053115 (2014).

\bibitem{ZhangNanotech2017}
Y.~Zhang, H.~Zhang, Y.~Cai, J.~Song, and P.~He,
\newblock Nanotechnology {\bfseries 28}, 075701 (2017).

\bibitem{BriggsNatMat2020}
N.~Briggs {\em et~al.},
\newblock Nature Materials {\bfseries 19}, 637 (2020).

\bibitem{SutterNatMat2008}
P.~W. Sutter, J.-I. Flege, and E.~A. Sutter,
\newblock Nature Materials {\bfseries 7}, 406 (2008).

\bibitem{McCartyCarbon2009}
K.~McCarty, P.~Feibelman, E.~Loginova, and N.~Bartelt,
\newblock Carbon {\bfseries 47}, 1806 (2009).

\bibitem{CuiPCCP2010}
Y.~Cui, Q.~Fu, and X.~Bao,
\newblock Physical Chemistry Chemical Physics {\bfseries 12}, 5053 (2010).

\bibitem{BorcaNJP2010}
B.~Borca {\em et~al.},
\newblock New Journal of Physics {\bfseries 12}, 093018 (2010).

\bibitem{HuangAPL2011}
L.~Huang {\em et~al.},
\newblock Applied Physics Letters {\bfseries 99}, 163107 (2011).

\bibitem{UlstrupSS2018}
S.~Ulstrup {\em et~al.},
\newblock Surface Science {\bfseries 678}, 57 (2018).

\bibitem{BauerBook2014}
E.~Bauer,
\newblock {\itshape Surface Microscopy with Low-Energy Electrons} (, 2014).

\bibitem{MoritzPRL2010}
W.~Moritz {\em et~al.},
\newblock Physical Review Letters {\bfseries 104}, 136102 (2010).

\bibitem{Bromberger2002}
C.~Bromberger, H.~Jänsch, and D.~Fick,
\newblock Surface Science {\bfseries 506}, 129 (2002).

\bibitem{ZangwillBook}
A.~Zangwill,
\newblock {\itshape Physics at {Surfaces}} (, 1988).

\bibitem{Topping1927}
J.~Topping,
\newblock Proc. R. Soc. Lond. {\bfseries 114}, 67 (1927).

\bibitem{VerhoefSS1997}
R.~W. Verhoef and M.~Asscher,
\newblock Surface Science {\bfseries 391}, 11 (1997).

\bibitem{Bundaleski2013}
N.~Bundaleski, J.~Trigueiro, A.~Silva, A.~Moutinho, and O.~Teodoro,
\newblock Journal of Applied Physics {\bfseries 113}, 183720 (2013),
  https://doi.org/10.1063/1.4804663.

\bibitem{MullikenJCP1934}
R.~Mulliken,
\newblock The Journal of Chemical Physics {\bfseries 2}, 782 (1934).

\bibitem{HinzeJACS1962}
J.~Hinze and H.~H. Jaffe,
\newblock Journal of the American Chemical Society {\bfseries 84}, 540 (1962).

\bibitem{Lee_Persson2012}
E.~Lee and K.~Persson,
\newblock Nano Letters {\bfseries 12}, 4624 (2012).

\bibitem{Dimakis2019}
N.~Dimakis {\em et~al.},
\newblock Molecules {\bfseries 24}, 754 (2019).

\bibitem{WandeltSS1981}
K.~Wandelt, J.~Hulse, and J.~Küppers,
\newblock Surface Science {\bfseries 104}, 212 (1981).

\bibitem{Raghavan2021}
A.~Raghavan {\em et~al.},
\newblock Phys. Chem. Chem. Phys. {\bfseries 23}, 7822 (2021).

\end{thebibliography}

\end{document}